\documentclass[
reprint,           
superscriptaddress,
amsmath,           
amssymb,           
prd,               
notitlepage,       
longbibliography,  
floatfix,          
twocolumn,
]{revtex4-1}

\usepackage{tensor}     
\usepackage{graphicx}   
\usepackage[
colorlinks=true,        
citecolor=blue,         
linkcolor=blue,         
urlcolor=blue           
]{hyperref}             
\usepackage{bm}         
\usepackage{xcolor}     

\newcommand{\newc}{\newcommand*} 
\newc{\figurewidth}{3.2in}
\newc{\xbar}{\bar{x}}
\newc{\rhoeq}{\rho_{\rm{eq}}}
\newc{\zeq}{z_{\rm{eq}}}
\newc{\la}{\lambda}
\newc{\tla}{\tilde{\la}}
\newc{\dt}{\delta}
\newc{\Dt}{\Delta}
\newc{\vj}{\vec{j}}
\newc{\vl}{\vec{l}}
\newc{\hx}{\hat{x}}
\newc{\hy}{\hat{y}}
\newc{\bj}{\bm{j}}
\newc{\mJ}{\mathcal{J}}
\newc{\mP}{\mathcal{P}}
\newc{\ga}{\gamma}
\newc{\Msun}{M_\odot}
\newc{\app}{\approx}
\newc{\av}[1]{\langle #1 \rangle}
\newc{\eq}[1]{Eq.~\eqref{#1}}
\newc{\al}{\alpha}
\newc{\Xstar}{X_{\ast}}
\newc{\seq}{\sigma_{\rm{eq}}}
\newc{\fpbh}{f_{\rm{pbh}}}
\newc{\VT}{\langle VT \rangle}

\def\p{\partial}

\def\({\left(}
\def\){\right)}
\def\[{\left[}
\def\]{\right]}

\def\e{\begin{equation}}
\def\q{\end{equation}}
\def\m{\begin{eqnarray}}
\def\n{\end{eqnarray}}

\begin{document}

\title{Stochastic gravitational-wave background from axion-monodromy oscillons in string theory during preheating}

\author{Yu Sang}
\email{sangyu@itp.ac.cn}
\affiliation{CAS Key Laboratory of Theoretical Physics, 
Institute of Theoretical Physics, Chinese Academy of Sciences,
Beijing 100190, China}
\author{Qing-Guo Huang}
\email{huangqg@itp.ac.cn}
\affiliation{CAS Key Laboratory of Theoretical Physics, 
Institute of Theoretical Physics, Chinese Academy of Sciences,
Beijing 100190, China}
\affiliation{School of Physical Sciences, 
University of Chinese Academy of Sciences, 
No. 19A Yuquan Road, Beijing 100049, China}
\affiliation{Center for Gravitation and Cosmology, College of Physical Science and Technology, Yangzhou University, Yangzhou 225009, China}
\affiliation{Synergetic Innovation Center for Quantum Effects 
and Applications, Hunan Normal University, Changsha 410081, China}

\date{\today}
\begin{abstract}
Axion monodromy is a generic phenomenon in string compactifications. We investigate the production of gravitational-wave background from axion monodromy inflation during preheating. By performing lattice simulations using pseudospectral methods, we find that significant single-peak stochastic gravitational-wave backgrounds are generated during preheating for asymptotic linear $|\phi|$ and $\phi^{2/3}$  potentials in axion monodromy inflation and cuspy potentials. This may allow us to explore string compactifications through gravitational waves.

\end{abstract}

\pacs{???}

\maketitle


\section{Introduction}
In string compactifications, axions arise from integrating gauge potentials over nontrivial cycles \cite{Halverson:2018xge,Marsh:2015xka,Baumann:2014nda}. The introduction of monodromy, a generic phenomenon in string compactifications, extends the field range of individual closed-string axions to super-Planck scale. A linear potential for the canonically normalized axion field is induced by axion monodromy introduced by space-filling wrapped five branes \cite{McAllister:2008hb}, and an asymptotic $\phi^{2/3}$  potential for the canonically normalized field $\phi$ is induced by using a monodromy of wrapped branes on nil manifolds which contain tori twisted over circles \cite{Silverstein:2008sg}. As a natural realization of large field chaotic inflation in string theory, axion monodromy model can still fit the constraints on tensor-to-scalar ratio and tilt of the scalar power spectrum using cosmic microwave background from Planck \cite{Aghanim:2018eyx}. 

It is interesting and worthy to probe the string compactifications by means of axion monodromy inflation. One way is to study  the possible phenomenological consequences of the dynamics in preheating after inflation, including production of oscillons and emission of gravitational waves. For example, see \cite{Lozanov:2019ylm,Ashoorioon:2013oha,Amin:2018xfe} etc. For the asymptotic linear potential case, the oscillons formation and gravitational-wave background production have been studied in analytical and numerical methods \cite{Amin:2011hj,Zhou:2013tsa}. If ignoring the small field region behaviors,  axion monodromy inflation leads to cuspy linear and fractional power potentials, in case of  which the productions of oscillons and gravitational-wave background are also investigated \cite{Liu:2017hua, Liu:2018rrt}. But in string axion monodromy, both the asymptotic linear $|\phi|$ and $\phi^{2/3}$ potentials are smooth around $\phi=0$ in \cite{McAllister:2008hb,Silverstein:2008sg}.

The goal of this paper is to study the gravitational-wave emission during preheating in axion monodromy inflation model via lattice simulation. Roughly speaking, two popular methods have been developed to simulate the dynamics of preheating. The first one is based on LATTICEEASY \cite{Felder:2000hq}, in which the ordinary differential equation of scalar field is solved using staggered leapfrog algorithm and the spatial derivatives of the scalar field are computed using finite differencing method, i.e. computing approximations to a derivative by subtracting neighboring-point values and then dividing by some multiple of the grid spacing. The other method is based on PSpectRe \cite{Easther:2010qz}, in which the equation of scalar field is solved using Fourier-space pseudospectral method. The potential and its derivatives are calculated by first converting the fields into their position-space representation, performing the necessary multiplications, and then taking the inverse transform. The spatial derivatives in pseudospectral methods are taken by multiplying the corresponding Fourier component by the wave number; therefore, the pseudospectral methods are free of differencing noise and provide more accurate results. It implies that PSpectRe will provide more accurate simulations than LATTICEEASY. 

In this paper, we simulate the production of gravitational waves during preheating from axion monodromy inflation. We perform lattice simulations using Fourier-space pseudospectral method to revisit the production of gravitational waves for the asymptotic linear potential and calculate the result for the asymptotic $\phi^{2/3}$  potential. We also adopt PSpectRe to calculate the  productions of gravitational waves for the cuspy potentials which were simulated using finite differencing method in \cite{Liu:2017hua, Liu:2018rrt} and find that the right peaks of gravitational-wave spectra given in \cite{Liu:2017hua, Liu:2018rrt} do not show up. The rest of this paper is organized as follows. In Sec. II, we introduce our method. Our results for the asymptotic $|\phi|$ and $\phi^{2/3}$  potentials are described in Secs. III and IV, respectively. Section V is the summary and discussion.

\section{Method}
The strong growth of field perturbations during preheating makes it a nonlinear process. We use lattice simulation to study the nonperturbative evolution of the inflaton field in an expanding universe. The following equations are solved on a discrete spacetime lattice:
\m
\ddot{\phi}\, + \, 3H\dot{\phi}\, - \,{1\over a^2}\nabla^2\phi\, + \, {\p  V \over \p \phi} \, = \, 0\,, \label{eq:EOM_fld} \\
H^2\, = \, {1\over 3M^2_{\rm{pl}}}\( V\,  + \, {1\over 2}\dot{\phi}^2\, +  {1\over 2a^2}\left|\nabla\phi\right|^2 \)\,, \label{eq:EOM_hubble}
\n
where $M_\mathrm{pl}=1/\sqrt{8\pi G}$ is the reduced Planck mass. The initial values of the inflaton field $\phi_i$ and its derivative $\dot{\phi}_i$ are determined by solving the  differential equations for the background evolution. The initial fluctuations of the inflaton field and its derivative are given by quantum vacuum fluctuations.

To calculate the gravitational-wave background during preheating, the evolution of the transverse and traceless (TT) part of metric perturbation is computed in the simulation. The equation of gravitational wave is given by 
\e
\ddot{h}_{ij}\, + \,3H\dot{h}_{ij}\, - \, {1\over a^2}\nabla^2h_{ij}\, = \,  {2\over M^2_{\rm{pl}} a^2}\Pi_{ij}~,
\label{eq:EOM_GW}
\q
where the $\Pi_{ij}$ is the TT part of the anisotropic stress of the scalar field
\e
\Pi_{ij}  =  \[\partial_i\phi\partial_j\phi\]^{\rm TT}\,.
\label{eq:effective_source_term}
\q
The energy density of the gravitational wave takes the form 
\e
\rho_{\rm GW}(t) = {M^2_{\rm pl} \over 4}\,\left\langle\dot{h}_{ij}(\textbf{x},t)\dot{h}_{ij}(\textbf{x},t)\right\rangle_{\rm{V}}\,,
\q
where $\left\langle...\right\rangle_{\rm{V}}$ denotes the spatially average over a sufficiently large volume. Finally, the spectrum of gravitational wave per logarithmic momentum interval is given by 
\e
\Omega_{\rm GW}(k) \, = \,{1 \over \rho_{\rm c}}\, {d\,\rho_{\rm GW} \over d \,{\rm ln} k}\,,
\q
where $\rho_{\rm c}$ is the critical density of the Universe. Provided that at the end of simulation, the Universe instantly reheats and all energy is immediately converted into radiation, we can convert the spectrum of gravitational wave to present observable values.

Our code is based on the publicly available PSpectRe \cite{Easther:2010qz}, which uses Fourier-space pseudospectral method to simulate the evolution of scalar field in an expanding universe. We extend PSpectRe to solve equation of  metric perturbation in Eq. (\ref{eq:EOM_GW}) using pseudospectral algorithm \cite{Easther:2007vj} and calculate the gravitational-wave spectrum. Both of the scalar field and metric perturbation evolutions are computed in fourth-order-in-time Runge-Kutta integration scheme. Our simulations are performed on a $256^3$ lattice.

\section{Asymptotic linear potential}
\begin{figure}[!htbp]
\centering
\includegraphics[width=.4\textwidth]{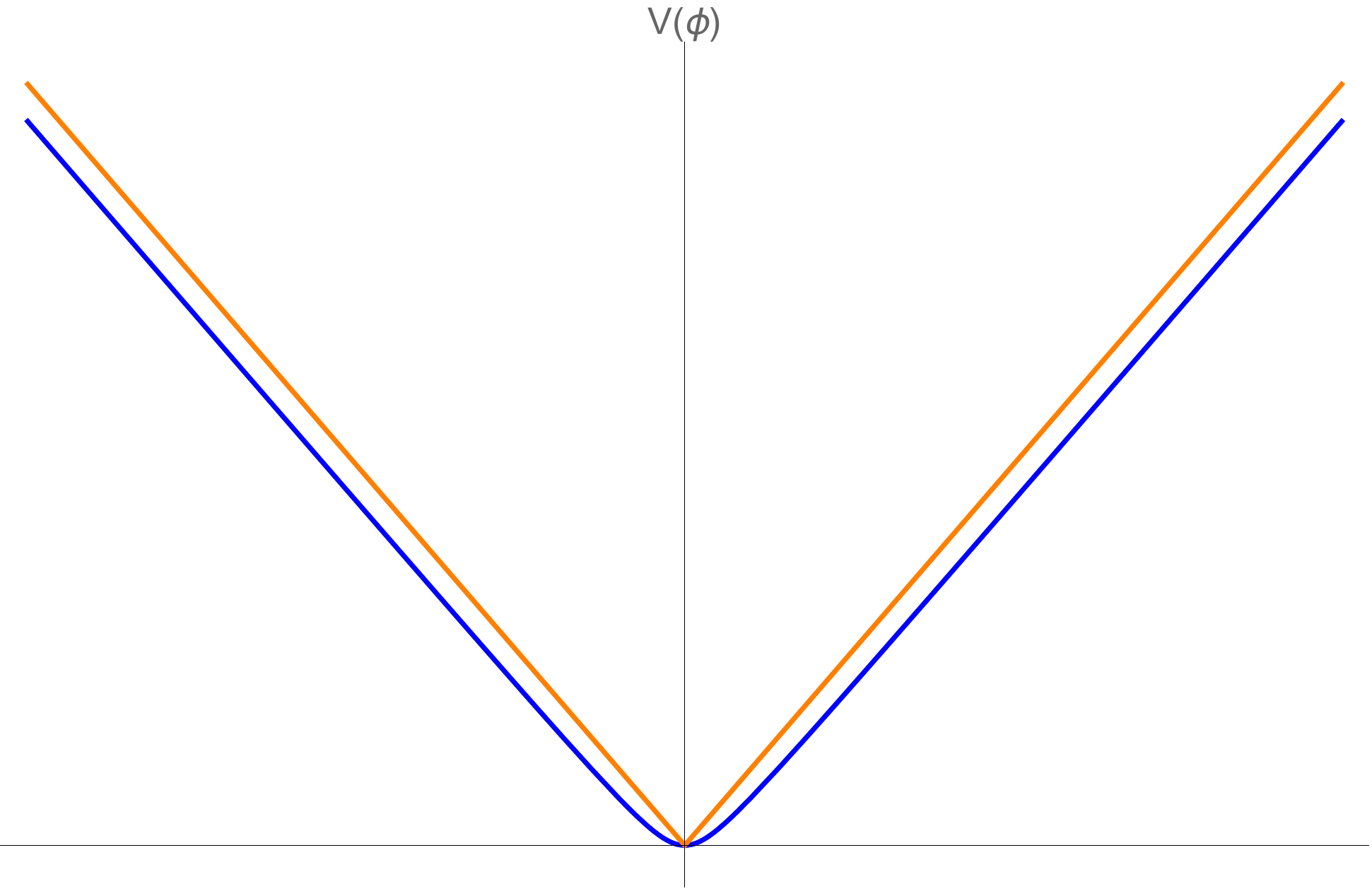}
\caption{Asymptotic linear potential in axion monodromy (blue) and cuspy linear potential (orange). }
\label{fig:V1}
\end{figure}

In axion monodromy model \cite{McAllister:2008hb},  the asymptotic linear potential is described by the following Lagrangian 
\m
{\cal L}(\phi)=-{(\p\phi)^2\over 2}-m^2M^2 \[\(1+{\phi^2\over M^2}\)^{1/2}-1\]. 
\label{eq:axionLphi1}
\n 
The potential term goes like $V(\phi) = m^2 \phi^2/2$ for $|\phi| /M\ll 1$, and $V(\phi) = m^2  M|\phi|$ for $|\phi| /M\gg 1$ (see Fig.~\ref{fig:V1}), satisfying the ``open up" condition of oscillon formation \cite{Amin:2010jq, Antusch:2017flz}. Self-resonance following inflation leads to copious oscillon generation and an oscillon-dominated phase \cite{Amin:2011hj}, and gravitational waves are significantly produced when oscillons are formed \cite{Zhou:2013tsa}.
In the limit of $M/M_\mathrm{pl} \ll 1$, the potential in Eq.~(\ref{eq:axionLphi1}) is reduced to a cuspy form,
\e
V(\phi)=\lambda_1 M_{\text{pl}}^3 |\phi|.
\label{eq:cuspyV1}
\q
In this section we consider the potentials in Eq.~(\ref{eq:axionLphi1}) and (\ref{eq:cuspyV1}), respectively.

In our lattice simulations, parameters in asymptotic linear potential are $m=1.0 \times 10^{-5} M_{\rm{pl}}$ and $M=0.01M_{\rm{pl}}$, and the coupling in cuspy linear potential is $\lambda_1 = 1.0 \times 10^{-12}$, and hence the potential terms in large field region are the same in these two models [axion monodromy in Eq.~(\ref{eq:axionLphi1}) and cuspy potential in Eq.~(\ref{eq:cuspyV1})]. The initial values of inflaton field and its derivative are determined to be  $\phi_i = 0.5 M_{\rm{pl}}$ and $\dot{\phi}_i =-6.0\times 10^{-7}M^2_{\rm{pl}}$ for both models, via evolving the background scaler field from slow-roll region. Since the present amplitude of gravitational waves is independent on the energy scale of inflation, and the present peak frequency is proportional to the  energy scale of inflation \cite{Easther:2006gt,Easther:2006vd}, we rescale the model parameters to match the sensitive frequency of LIGO.

\begin{figure}[!htbp]
\centering
\includegraphics[width=.4\textwidth]{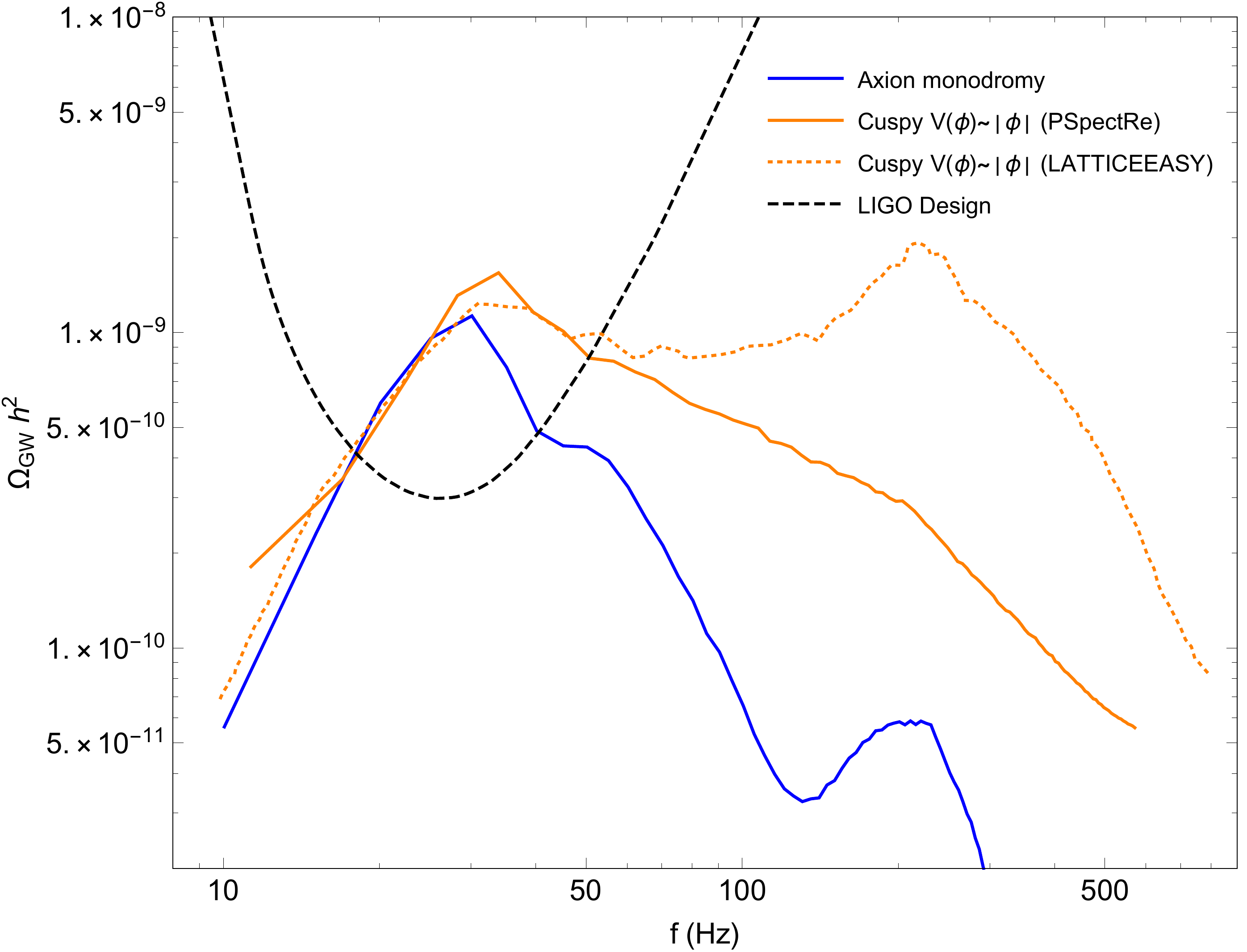}
\caption{Present gravitational-wave spectra for string axion monodromy with asymptotic linear potential for $m=  1.14\times 10^{-19} M_{\rm{pl}}$ and $M=0.01M_{\rm{pl}}$ (blue solid curve), and cuspy potential with potential $V\sim |\phi|$ for $\lambda_1  = 1.3\times 10^{-40}$. The orange dotted curve is calculated by LATTICEEASY given in \cite{Liu:2017hua}, and the orange solid curve is calculated by PSpectRe.}
\label{fig:gw_1}
\end{figure}

Our results are illustrated in Fig.~\ref{fig:gw_1}. The blue solid curve is the present gravitational-wave spectrum for the asymptotic linear potential in axion monodromy model with parameter $m$ rescaled to $m=  1.14\times 10^{-19} M_{\rm{pl}}$. The orange solid curve is the present gravitational-wave spectrum for the cuspy linear potential with parameter $\lambda_1 $ rescaled to $\lambda_1  = 1.3\times 10^{-40}$. Comparing to the result (orange dotted curve) in \cite{Liu:2017hua} with the same parameters, but  simulated using LATTICEEASY, our result roughly recovers the left peak in \cite{Liu:2017hua}, but there is no right peak in our simulation by adopting PSpectRe. Actually, the right peak in \cite{Liu:2017hua} is induced by the large-k modes, or equivalently the fluctuations in the small scales, and it might be sensitive to the differencing noise. In principle, the PSpectRe is supposed to be free of differencing noise and provide more accurate results.

\section{Asymptotic $\phi^{2/3}$ potential}

\begin{figure}[!htbp]
\centering
\includegraphics[width=.4\textwidth]{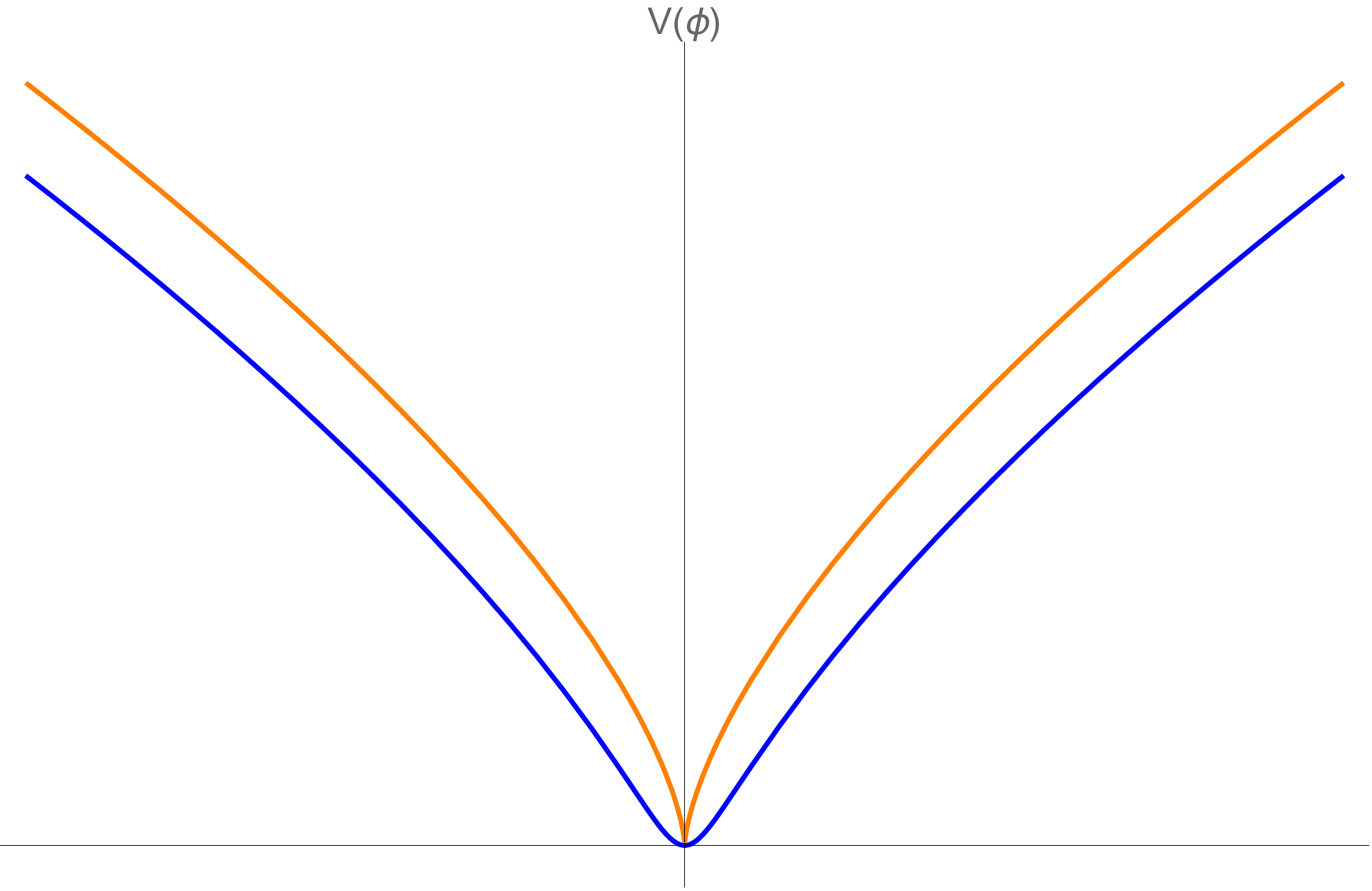}
\caption{Asymptotic $\phi^{2/3}$ potential in axion monodromy (blue) and cuspy $\phi^{2/3}$ potential (orange).}
\label{fig:V23}
\end{figure}

The asymptotic $\phi^{2/3}$ potential in axion monodromy model \cite{Silverstein:2008sg} is associated with a Lagrangian as follows 
\begin{align}
\nonumber
{\cal L}(\psi)=&-\sqrt{1+\({\psi\over \psi_c}\)^2} {(\p\psi)^2\over 2}\\
\label{eq:axionLpsi}
&-\lambda M_{\text{pl}}^4 \(\sqrt{1+\({\psi\over \psi_c}\)^2}-1\). 
\end{align}
It can be converted into a canonical form 
\m
{\cal L}(\phi)=-{(\p\phi)^2\over 2}-V(\phi)
\n 
by introducing a canonical field $\phi$ with
\m
d\phi=\[1+\({\psi\over \psi_c}\)^2 \]^{1/4} d\psi. 
\label{eq:axionLphi23}
\n 
The potential goes like $V(\phi) =\lambda M_{\text{pl}}^4 \phi^2/2 \psi_c^2 $ for $|\psi| / \psi_c\ll 1$, and $V(\phi) = \lambda M_{\text{pl}}^4 ( 3\phi / 2 \psi_c )^{2/3}$ for $|\psi|/ \psi_c\gg 1$ (see Fig.~\ref{fig:V23}). This potential satisfies the ``open up" condition of oscillon formation \cite{Amin:2010jq, Antusch:2017flz}. Hence, the oscillons are expected to be generated copiously and seed a significant gravitational-wave background. 
In the canonical frame, one can see that $\phi(|\psi|= \sqrt{2}\psi_c)$ is the inflection points of potential $V(\phi)$, or equivalently $d^2V/d\phi^2 <0$ when $|\psi|> \sqrt{2}\psi_c$ and $d^2V/d\phi^2 >0$ when $|\psi|< \sqrt{2}\psi_c$.
Different from model with potential (\ref{eq:axionLphi1}) in which the parametric resonance leads to the growth of fluctuations and hence formation of oscillons, the tachyonic preheating and tachyonic oscillations provide other mechanisms for generating large inhomogeneities in model with potential (\ref{eq:axionLpsi}).  
In limit of $\psi_c/M_{\text{pl}}\ll 1$, the potential in model (\ref{eq:axionLpsi}) is reduced to a cuspy form as follows:
\e
V(\phi)=\lambda_{2} M_{\text{pl}}^{10/3} \phi^{2/3}.
\label{eq:cuspyV23}
\q
Such a cusp may also trigger copious oscillon formation and gravitational-wave production \cite{Liu:2017hua, Liu:2018rrt}. 

In our lattice simulations, parameters in the asymptotic $\phi^{2/3}$ potential are $\lambda = 3.54 \times 10^{-14} $ and $\psi_c=0.01M_{\rm{pl}}$, and the coupling in cuspy $\phi^{2/3}$ potential is $\lambda_{2} = 1.0 \times 10^{-12}$, such that potential terms in large field region are the same in these two models. The initial values of inflaton field and its derivative are determined to be  $\phi_i = 0.5 M_{\rm{pl}}$ and $\dot{\phi}_i =-4.5\times 10^{-7}M^2_{\rm{pl}}$ for both models by evolving the background scaler field from slow-roll region. Similar to the case of asymptotic linear potential, we also rescale the model parameters to match the sensitive frequency of LIGO. 

\begin{figure}[!htbp]
\centering
\includegraphics[width=.4\textwidth]{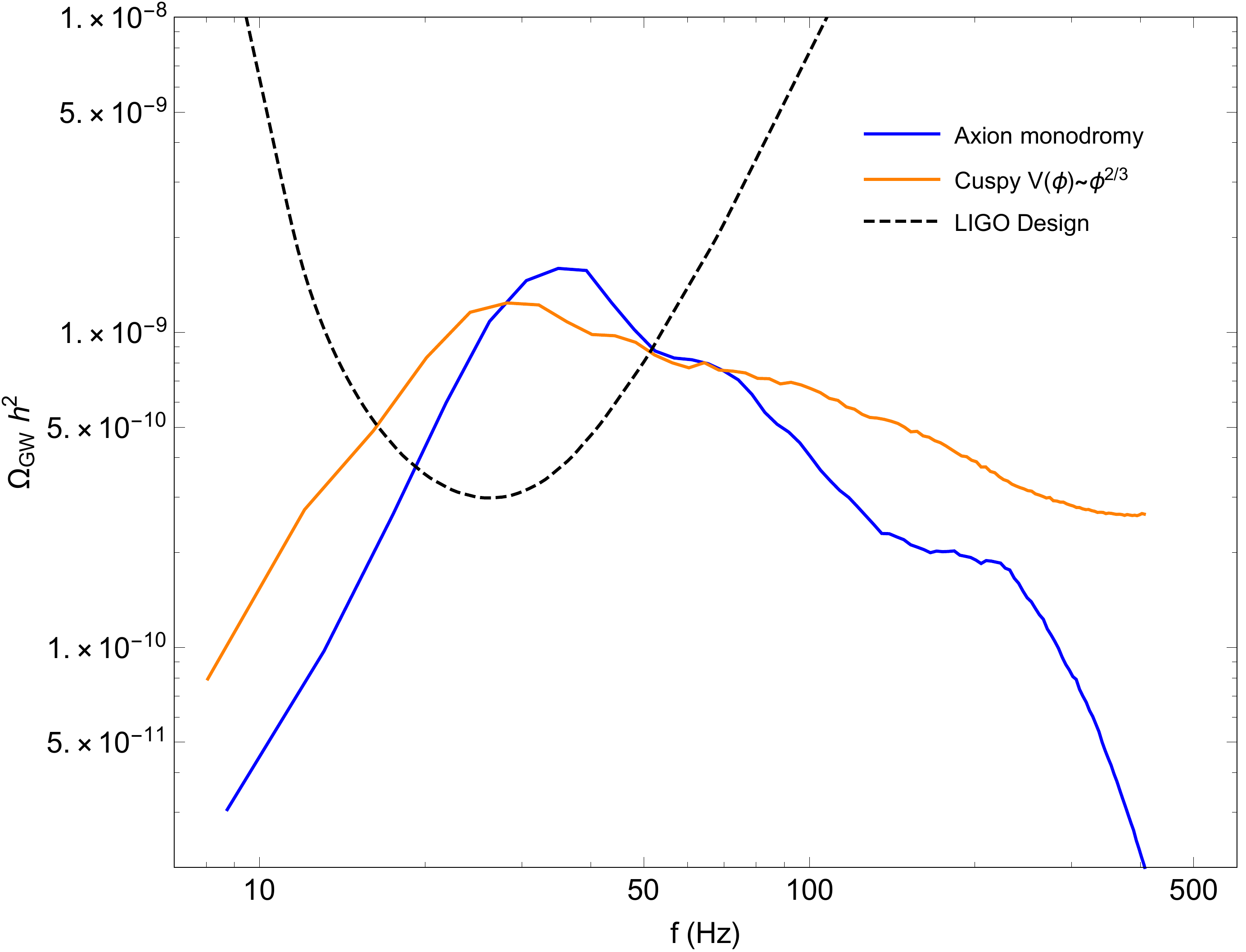}
\caption{Present gravitational-wave spectra for string axion monodromy with asymptotic $\phi^{2/3}$ potential for $\lambda = 3.54 \times 10^{-42} $ and $\psi_c=0.01M_{\rm{pl}}$ (blue solid curve), and cuspy potential with potential $V\sim \phi^{2/3}$ for $\lambda_{2} = 1.0 \times 10^{-40}$ (orange solid curve).}
\label{fig:gw_23}
\end{figure}

Our results are illustrated in Fig. \ref{fig:gw_23}. The blue solid curve is the present gravitational-wave spectrum for the asymptotic $\phi^{2/3}$ potential in axion monodromy model with parameter $\lambda$ rescaled to $\lambda = 3.54 \times 10^{-42}$. 
The orange solid curve corresponds to the present gravitational-wave spectrum for the cuspy $\phi^{2/3}$ potential with parameter $\lambda_2 $ rescaled to $\lambda_2  = 1.0\times 10^{-40}$. Again there are only one significant peak in the gravitational-wave spectrum for both the axion monodromy potential and cuspy potential.

\section{Summary and Discussion}

To summarize, we simulate the production of stochastic gravitational-wave background from axion monodromy inflation during preheating based on pseudospectral algorithm. We find significant productions of stochastic gravitational-wave background with single peak during preheating for the asymptotic $|\phi|$ and $\phi^{2/3}$ potentials in axion monodromy inflation and cuspy potentials. Different from most of works simulating scalar field evolution and gravitational-wave production during preheating in which the finite differencing method was used, our simulations use Fourier-space pseudospectral method which is free of differencing noise and has excellent spatial resolution. Results from finite differencing method and pseudospectral method have similar peak frequency and spectrum magnitude, but the details of peak structure are different. Different from the results simulated by using finite differencing method for the cuspy potentials in \cite{Liu:2017hua, Liu:2018rrt}, we obtain significant single-peak gravitational-wave spectra by adopting pseudospectral method for string axion monodromy potentials and cuspy potentials. 

Actually, the method employed in our work is more accurate than that used in \cite{Liu:2017hua}. The time integration in \cite{Liu:2017hua} is employed in second order staggered leapfrog method and the error in time integration is of order $(\Delta t)^3$, where $\Delta t$ is the time step, while we used fourth order Runge-Kutta method and hence the error in time integration is of order $(\Delta t)^5$. Second, the authors used the finite differencing method to calculate the spatial derivatives in \cite{Liu:2017hua},  while we used pseudospectral method. Differencing noise inevitably arises in finite differencing method, but is avoided in pseudospectral method, which implies that pseudospectral method is more accurate than finite differencing method. For example, as shown in \cite{Easther:2010qz}, pseudospectral method can achieve reliable results with a $64^3$ grid in circumstances where finite differencing method requires $128^3$ or even $256^3$ points.

\section*{Acknowledgments}

We would like to thank Jing Liu and Shuang-Yong Zhou for helpful discussion. 
We acknowledge the use of HPC Cluster of ITP-CAS. 
Y. S. is supported by grants from NSFC (Grant No. 11847218). 
Q. G. H is supported by grants from NSFC 
(Grants No. 11690021, No. 11575271, and No. 11747601), 
the Strategic Priority Research Program of Chinese Academy of Sciences 
(Grants No. XDB23000000 and No. XDA15020701), and Top-Notch Young Talents Program of China.




\begin{thebibliography}{99}
\frenchspacing

    
\bibitem{Baumann:2014nda} 
  D.~Baumann and L.~McAllister,
  arXiv:1404.2601 [hep-th].
  
    
\bibitem{Marsh:2015xka} 
  D.~J.~E.~Marsh,
  Phys.\ Rept.\  {\bf 643}, 1 (2016)
  [arXiv:1510.07633 [astro-ph.CO]].

\bibitem{Halverson:2018xge} 
  J.~Halverson and P.~Langacker,
  PoS TASI {\bf 2017}, 019 (2018)
  [arXiv:1801.03503 [hep-th]].  
  
\bibitem{McAllister:2008hb} 
  L.~McAllister, E.~Silverstein and A.~Westphal,
  Phys.\ Rev.\ D {\bf 82}, 046003 (2010)
  [arXiv:0808.0706 [hep-th]].

\bibitem{Silverstein:2008sg} 
  E.~Silverstein and A.~Westphal,
  Phys.\ Rev.\ D {\bf 78}, 106003 (2008)
  [arXiv:0803.3085 [hep-th]].
  
\bibitem{Aghanim:2018eyx} 
  N.~Aghanim {\it et al.} [Planck Collaboration],
  arXiv:1807.06209 [astro-ph.CO].
  
\bibitem{Lozanov:2019ylm} 
  K.~D.~Lozanov and M.~A.~Amin,
  Phys.\ Rev.\ D {\bf 99}, no. 12, 123504 (2019)
  [arXiv:1902.06736 [astro-ph.CO]].  
  
\bibitem{Ashoorioon:2013oha} 
  A.~Ashoorioon, B.~Fung, R.~B.~Mann, M.~Oltean and M.~M.~Sheikh-Jabbari,
  JCAP {\bf 1403}, 020 (2014)
  [arXiv:1312.2284 [hep-th]].
  
  \bibitem{Amin:2018xfe} 
  M.~A.~Amin, J.~Braden, E.~J.~Copeland, J.~T.~Giblin, C.~Solorio, Z.~J.~Weiner and S.~Y.~Zhou,
  Phys.\ Rev.\ D {\bf 98}, 024040 (2018)
  [arXiv:1803.08047 [astro-ph.CO]].
  
  
  \bibitem{Amin:2011hj} 
  M.~A.~Amin, R.~Easther, H.~Finkel, R.~Flauger and M.~P.~Hertzberg,
  Phys.\ Rev.\ Lett.\  {\bf 108}, 241302 (2012)
  [arXiv:1106.3335 [astro-ph.CO]].
  
\bibitem{Zhou:2013tsa} 
  S.~Y.~Zhou, E.~J.~Copeland, R.~Easther, H.~Finkel, Z.~G.~Mou and P.~M.~Saffin,
  JHEP {\bf 1310}, 026 (2013)
  [arXiv:1304.6094 [astro-ph.CO]].
  
\bibitem{Liu:2017hua} 
  J.~Liu, Z.~K.~Guo, R.~G.~Cai and G.~Shiu,
  Phys.\ Rev.\ Lett.\  {\bf 120}, no. 3, 031301 (2018)
  [arXiv:1707.09841 [astro-ph.CO]].
  
\bibitem{Liu:2018rrt} 
  J.~Liu, Z.~K.~Guo, R.~G.~Cai and G.~Shiu,
  Phys.\ Rev.\ D {\bf 99}, no. 10, 103506 (2019)
  [arXiv:1812.09235 [astro-ph.CO]].
    
\bibitem{Felder:2000hq} 
  G.~N.~Felder and I.~Tkachev,
  Comput.\ Phys.\ Commun.\  {\bf 178}, 929 (2008)
  [hep-ph/0011159].
  
\bibitem{Easther:2010qz} 
  R.~Easther, H.~Finkel and N.~Roth,
  JCAP {\bf 1010}, 025 (2010)
  [arXiv:1005.1921 [astro-ph.CO]].
  

  
\bibitem{Easther:2007vj} 
  R.~Easther, J.~T.~Giblin and E.~A.~Lim,
  Phys.\ Rev.\ D {\bf 77}, 103519 (2008)
  [arXiv:0712.2991 [astro-ph]].
  
\bibitem{Amin:2010jq} 
  M.~A.~Amin and D.~Shirokoff,
  Phys.\ Rev.\ D {\bf 81}, 085045 (2010)
  [arXiv:1002.3380 [astro-ph.CO]].

\bibitem{Antusch:2017flz} 
  S.~Antusch, F.~Cefala, S.~Krippendorf, F.~Muia, S.~Orani and F.~Quevedo,
  JHEP {\bf 1801}, 083 (2018)
  [arXiv:1708.08922 [hep-th]].

\bibitem{Easther:2006vd} 
  R.~Easther, J.~T.~Giblin, Jr. and E.~A.~Lim,
  Phys.\ Rev.\ Lett.\  {\bf 99}, 221301 (2007)
  [astro-ph/0612294].
  
\bibitem{Easther:2006gt} 
  R.~Easther and E.~A.~Lim,
  JCAP {\bf 0604}, 010 (2006)
  [astro-ph/0601617].
  


\end{thebibliography}
\end{document}